# Preparation and superconductivity of iron selenide thin films

Y Han, W Y Li, L X Cao[1], S Zhang, B Xu and B R Zhao

National Laboratory for Superconductivity, Institute of Physics and Beijing National Laboratory for Condensed

Matter Physics,

Chinese Academy of Sciences, Beijing 100190, China

Email: lxcao@aphy.iphy.ac.cn

**Abstract**

FeSe$_x$ (x = 0.80, 0.84, 0.88, 0.92) thin films were prepared on SrTiO$_3$(001) (STO), (La,Sr)(Al,Ta)O$_3$(001)

(LSAT), and LaAlO$_3$(001) (LAO) substrates by pulsed laser deposition method. All thin films show single-phase and c-axis oriented epitaxial growth, and are superconducting. Among them, the FeSe$_{0.88}$ thin films show T$_{c, onset}$ of 11.8 K and T$_{c, 0}$ of 3.4 K. The upper critical magnetic field is estimated to be 14.0 T.

---

[1] Author to whom any correspondence should be addressed.

Since the discovery of $LaFeAsO_{1-x}F_x$ superconductor [1], progress has been made very rapidly and four kinds of iron-based superconductors in total have been found: [1-10] (1) F-doped ROFeAs quarternary compound (1111 phase, R = La, Ce, Pr, Nd, Sm, Gd, etc) [1-5], (2) double FeAs layered $AFe_2As_2$ ternary compound (122 phase, A = Ca, Sr, Ba, with or without alkali metal doping ) [6,7], (3) single FeAs layered LiFeAs ternary compound (111 phase) [8,9], and (4) α-FeSe binary compound (11 phase) [10]. Among them, α-FeSe is very interesting since it possesses the simplest crystal structure, especially, it is the arsenic-free compound and therefore less toxic in nature in comparison with other three type of pnictide compounds. For this kind of superconductors, the superconducting transition temperature can be enhanced up to 27 K / 13.5 K ($T_{c, onset}$ / $T_{c,0}$) under high pressure [11], and 15.3 K / 11.8 K ($T_{c, onset}$ / $T_{c,0}$) via Te substitution [12-13]. In respect of experimental research on intrinsic physical properties [14-16] and device applications [17], high quality epitaxial thin films are necessary to be developed. Therefore attempts have been made to get epitaxial and superconducting arsenide thin films [18-20]. For example, the F-doped LaOFeAs thin films have been made and show an onset of superconductivity of 11 K [18], while for LaOFeAs thin film without F doping no superconducting transition was observed [19]. Hiramatsu et al. [20] had fabricated Co-doped $SrFe_2As_2$ thin films which showed $T_{c, onset}$ of 20 K and complete superconducting transition at ~15 K. It should be indicated that thin films in all these studies regarding 1111 phase and 122 phase are nearly phase pure epitaxy with minor impurity phases inside [18-20]. In this paper, we report the synthesis of iron selenide superconductor $FeSe_x$ (x = 0.80, 0.84, 0.88, 0.92) thin films. By pulsed laser deposition, we obtained single-phased, c-axis oriented epitaxial films. The films show $T_{c, onset}$ of 11.8 K and $T_{c, 0}$ of 3.4 K with x = 0.88. The upper critical field is estimated to be 14.0 T.

The thin films were pulsed laser deposited in a vacuum chamber with base pressure better than $4 \times 10^{-5}$ Pa. XeCl excimer laser (308 nm wavelength) with repetition rate of 2 Hz and power density of 100 mJ/mm$^2$ was irradiated on a polycrystalline FeSe$_x$ (x = 0.80, 0.84, 0.88, 0.92) target, which was sintered twice at 600 ˚C for 24 hours in vacuum sealed quartz tubes after the mixture of Fe and Se powders were thoroughly ground. The X-ray diffraction (XRD) pattern of one such prepared target (x = 0.88) and also its R-T and M-T measurement results (x = 0.88) are reproduced in figure 1(a) and figure 2(a), respectively. It can be seen that the main synthesis product is tetragonal α-FeSe with minor impurity hexagonal β-FeSe indicated as a star symbol for its strongest diffraction line, c.f. figure 1(a). From its R-T measurement, the zero resistance temperature is 5.2 K and onset temperature is 14.4 K; from its M-T measurement, the transition temperature is 7.9 K, as shown in figure 2(a).

It is worthy to note that plotting the XRD intensity in logarithmic scale is common to identify and to recognize the weaker diffraction peaks, in comparison with plotting in linear scale. The humps of about 200 cps in intensity in small diffraction angles (10 − 30 degree) are artifacts coming from the diffractometer.

The deposition procedure and parameters were studied and optimized systematically for all 4 nominal compositions, namely, x = 0.80, 0.84, 0.88, and 0.92, for over 150 thin film samples on 3 type of substrates, i.e., SrTiO$_3$(001) (STO), (La,Sr)(Al,Ta)O$_3$(001) (LSAT), and LaAlO$_3$(001) (LAO). The films given in this paper are all about 200 nm thick.

In general, it was found that the pure single-phased, c-axis oriented $FeSe_x$ films can be obtained on all three type of substrates, as given in figure 1(b) and (c) for those on STO and LSAT, respectively. The deposition parameters of these 2 films are the same as those for the best films grown on LAO (c.f. figure 1(f)), the details of which will be given later. Since the lattice mismatch is the smallest for $FeSe_x$ films on LAO and also since the physical properties are much better for films on LAO than on STO and LSAT, we concentrate on the LAO cases.

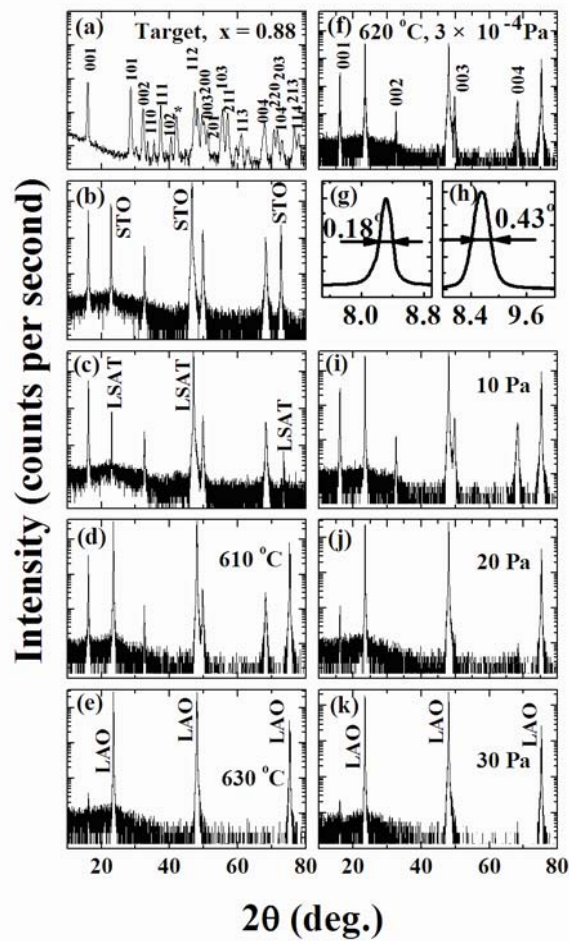

Figure 1. X-ray diffraction spectra of $FeSe_{0.88}$ (a) target, (b) film on STO at 620 ˚C under $3\times10^{-4}$ Pa, (c) film on LSAT at 620 ˚C under $3\times10^{-4}$ Pa, (d) film on LAO at 610 ˚C under $3\times10^{-4}$ Pa, (e) film on LAO at 630 ˚C under $3\times10^{-4}$ Pa, (f) film on LAO at 620 ˚C under $3\times10^{-4}$ Pa, (i) film on LAO at 620 ˚C under 10 Pa, (j) film on LAO at 620 ˚C under 20 Pa, and (k) film

on LAO at 620 ˚C under 30 Pa. Rocking curves of (001) peak of films given in (f) and (i) are reproduced in (g) and (h), respectively.

For films with x = 0.88 deposited on LAO under different conditions, some of their XRD patterns are given in figure 1(d)-(k) as examples. From crystal structure point of view, the pure single-phased, c-axis oriented FeSe$_x$ films can be obtained in a relatively large threshold of deposition pressure and temperature, e.g. from several tenths Pascal down to vacuum deposition, and from 630 ˚C down to 500 ˚C, c.f., figure 2(b) as a phase diagram. It is obvious that the films deposited at 620 ˚C under $3\times10^{-4}$ Pa, i.e., vacuum deposition without argon gas intentionally lead into the chamber, show best crystalline quality among all samples studied (c.f. figure 1(f)). For example, the full widths of half maximum of the rocking curves of FeSe$_{0.88}$(001) diffraction peak are 0.18˚ and 0.43˚ (c.f. figure 1(g) and (h)), respectively, for films deposited under vacuum (c.f. figure 1(f)) and 10 Pa (c.f. figure 1(i)), both at 620 ˚C. It is interesting to note that usually some kind of gases are intentionally introduced into the deposition chamber in order to adjust, to control the plume of the pulsed laser deposition, in many studies Ar. In this study, it was found that with or without Ar do influence the quality of the films, which may call further studies concerning the detailed information on microstructures and/or homogeneities of the structural and electronic phase(s).

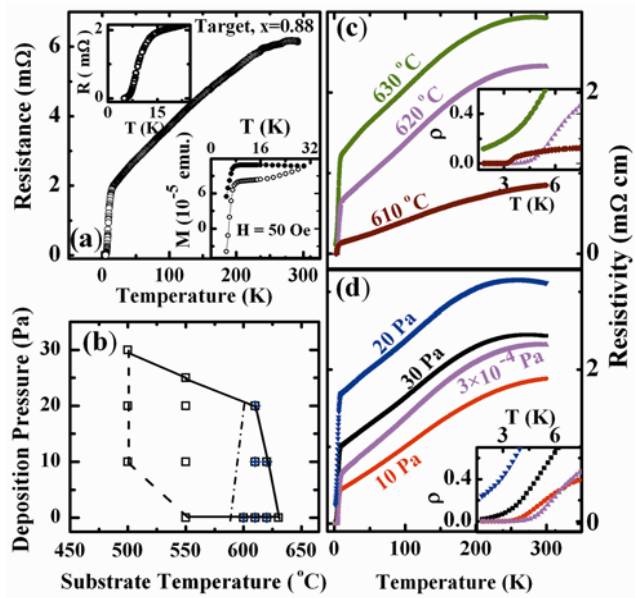

Figure 2. (a) Dependency of resistance and magnetization (lower right inset) on temperature of the FeSe$_{0.88}$ target pellet, with the upper left inset as the enlargement of R-T curve. (b) Phase diagram of FeSe$_{0.88}$ film deposition, with empty square labeled by which pure single-phased epitaxial films could be obtained, and cross symbol superconductivity detected. (c) Resistivity versus temperature for FeSe$_{0.88}$ film on LAO deposited under 3×10$^{-4}$ Pa at different temperature. (d) Resistivity v.s. temperature for FeSe$_{0.88}$ film on LAO deposited at 620 ˚C under different pressure.

However, the physical properties of the films are not that tolerant of deposition condition in comparison with their crystal structure. Thin films with good superconducting transitions can only be obtained in a relatively small area on the phase diagram as shown in figure 2(b). Given in figure 2(c) and (d), the best performance of superconductivity appears at around 620 ˚C and 3×10$^{-4}$ Pa, with T$_{c,}$ $_{onset}$ / T$_{c,0}$ of 11.8 K / 3.4 K. This is in accordance with the above crystal structure studies (c.f. figure 1(f) and (g)).

The surface morphologies of the films shown in figure 1(d)-(f) and figure 2(c) were observed by scanning electron microscope (SEM). Reproduced in figure 3, it is a bit surprise that the morphologies are quite different although all other parameters are the same, including composition, deposition pressure, cooling procedure, etc, with only minor changes in deposition temperature. The film deposited at 610 ˚C under $3\times10^{-4}$ Pa is quite smooth showing a featureless surface, which suggests Frank-van der Merwe growth mode (c.f. figure 3(a)); while the film deposited at 630 ˚C Stranski-Krastanov growth mode (c.f. figure 3(c)). For film deposited at 620 ˚C, which shows best qualities in both structure and superconductivity, its surface feature just lies in between the ones at 610 ˚C and 630 ˚C. It is either a Frank-van der Merwe growth with nanometer-scale dots precipitated on the surface or an early stage of a Stranski-Krastanov growth (c.f. figure 3(b) and (d)). However, it possesses microcracks which are ~1 μm away from each other and are aligned along the principle axes of LAO, i.e., [100] or [010]. This may result from the difference of crystalline structures between the LAO and $FeSe_x$, the former of which is pseudo-cubic and the later tetragonal. The minor lattice distortion originated from the shear strain determines the critical thickness and the film microstructures [21]. Furthermore, the phase transitions for LAO at ~435 ˚C and for α-FeSe at 457 ˚C may also influence the microstructure and morphology of the films.

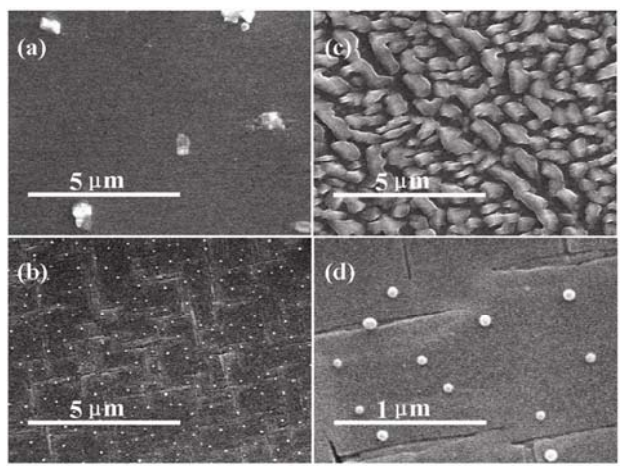

Figure 3. SEM pictures of $FeSe_{0.88}$ films on LAO(001) deposited under $3\times10^{-4}$ Pa at (a) 610 ˚C, (b) 620 ˚C, and (c) 630

°C. (d) Same surface as of (b) with higher magnification.

The superconducting feature of the FeSe$_x$ thin films with x = 0.80, 0.84, 0.88, and 0.92 are shown in figure 4(a), in which the superconducting transition temperatures are summarized. It looks to have a dome-like feature with the highest critical temperature at x = 0.88. Therefore, we further make study on influence of magnetic field on superconductivity of FeSe$_{0.88}$ thin films which are deposited at 620 °C under 3×10$^{-4}$ Pa. For this fact, the magnetic fields up to 5 T are applied along c-axis, and the R-T(H) data are shown in figure 4(b). The transition curve shifted to lower temperature while the field increased. As shown in figure 4(b), it can be seen that similar to the cuprate superconductor, the onset superconducting transitions are the same for all testing magnetic fields, and the broadening of superconducting transition occurs with increasing magnetic field. This implies the similar mode of vortex motion as in cuprate superconductor. By using the middle point of superconducting transition temperature to determine T$_c$ for each field, the upper critical field H$_{c,2}$(0) can be estimated to be 14.0 T according to WHH model [22], as shown in the inset of figure 4(b). This value is close to that obtained from the polycrystalline samples [10].

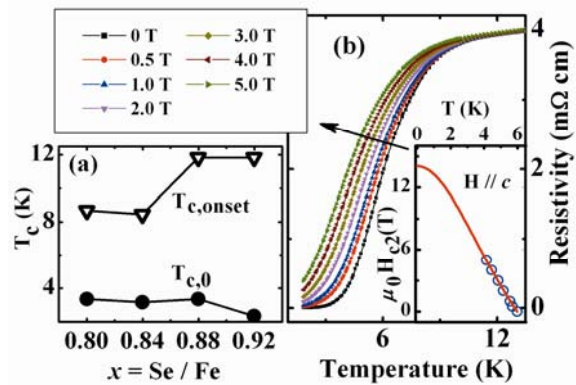

Figure 4. (a) Dependency of T$_{c,onset}$ and T$_{c,0}$ on Se content for films deposited on LAO at 620 °C under 3×10$^{-4}$ Pa. (b) Resistivity versus temperature in the external magnetic field. The zero-temperature upper critical field is ~14.0 T, as given in the inset of panel (b).

In summary, we deposited pure iron-based superconducting thin films, in particular, c-oriented iron selenide $FeSe_x$ films on STO, LSAT, and LAO, with x = 0.80, 0.84, 0.88, and 0.92. For x = 0.88, the films on LAO show $T_{c,\,onset}$ / $T_{c,0}$ of 11.8 K / 3.4 K, and $H_{c2}$ of 14.0 T.

This work is supported by the MOST of China (2006CB921107 and 2004CB619004), the NSFC (10774165), and the BRJH of the CAS.

Figure Captions:

Figure 1.    X-ray diffraction spectra of FeSe$_{0.88}$ (a) target, (b) film on STO at 620 ˚C under $3\times10^{-4}$ Pa, (c) film on LSAT at 620 ˚C under $3\times10^{-4}$ Pa, (d) film on LAO at 610 ˚C under $3\times10^{-4}$ Pa, (e) film on LAO at 630 ˚C under $3\times10^{-4}$ Pa, (f) film on LAO at 620 ˚C under $3\times10^{-4}$ Pa, (i) film on LAO at 620 ˚C under 10 Pa, (j) film on LAO at 620 ˚C under 20 Pa, and (k) film on LAO at 620 ˚C under 30 Pa. Rocking curves of (001) peak of films given in (f) and (i) are reproduced in (g) and (h), respectively.

Figure 2.    (a) Dependency of resistance and magnetization (lower right inset) on temperature of the FeSe$_{0.88}$ target pellet, with the upper left inset as the enlargement of R-T curve. (b) Phase diagram of FeSe$_{0.88}$ film deposition, with empty square labeled by which pure single-phased epitaxial films could be obtained, and cross symbol superconductivity detected. (c) Resistivity versus temperature for FeSe$_{0.88}$ film on LAO deposited under $3\times10^{-4}$ Pa at different temperature. (d) Resistivity v.s. temperature for FeSe$_{0.88}$ film on LAO deposited at 620 ˚C under different pressure.

Figure 3.    SEM pictures of FeSe$_{0.88}$ films on LAO(001) deposited under $3\times10^{-4}$ Pa at (a) 610 ˚C, (b) 620 ˚C, and (c) 630 ˚C. (d) Same surface as of (b) with higher magnification.

Figure 4.    (a) Dependency of T$_{c,onset}$ and T$_{c,0}$ on Se content for films deposited on LAO at 620 ˚C under $3\times10^{-4}$ Pa. (b) Resistivity versus temperature in the external magnetic field. The zero-temperature upper critical field is ~14.0 T, as given in the inset of panel (b).